# Superconductivity in Ternary Scandium Telluride Sc$_6$$M$Te$_2$ with 3$d$, 4$d$, and 5$d$ Transition Metals


Yusaku Shinoda[1], Yoshihiko Okamoto[2*], Youichi Yamakawa[3], Haruka Matsumoto[2], Daigorou Hirai[1], Koshi Takenaka[1]

[1]*Department of Applied Physics, Nagoya University, Nagoya 464-8603, Japan*
[2]*Institute for Solid State Physics, University of Tokyo, Kashiwa 277-8581, Japan*
[3]*Department of Physics, Nagoya University, Nagoya 464-8602, Japan*



We report the discovery of bulk superconductivity in Sc$_6$$M$Te$_2$ with seven kinds of transition-metal elements $M$. The critical temperatures for $M$ = 3$d$ elements are higher than those for 4$d$ and 5$d$ elements and increase in the order of $M$ = Ni, Co, and Fe with the highest $T_c$ of 4.7 K in Sc$_6$FeTe$_2$. First principles calculations indicate the presence of significant contribution of Fe 3$d$ orbitals at the Fermi energy, which most likely enhance the $T_c$ of Sc$_6$FeTe$_2$. The upper critical field for $M$ = Os is considerably enhanced by the strong spin-orbit coupling. These results show Sc$_6$$M$Te$_2$ to constitute a unique family of $d$-electron superconductors, in which $d$ electrons of 3$d$ and 5$d$ $M$ atoms strongly influence the superconducting properties.


Superconductivity in $d$-electron systems has been one of the central issues in condensed matter physics. There are many outstanding superconductors, such as cuprates[1], iron arsenides and selenides[2,3], spinel LiTi$_2$O$_4$[4], layered perovskite Sr$_2$RuO$_4$[5], hydrate Na$_x$CoO$_2$·$n$H$_2$O[6], β-pyrochlore KOs$_2$O$_6$[7], and kagome CsV$_3$Sb$_5$[8]. The $d$ electrons of the transition metals play a major role in determining their electrical properties and the cooperation between various features of $d$ electrons, such as strong electron correlation, strong spin–orbit coupling, and strong spin and orbital fluctuations, results in the emergence of unique unconventional superconductivities. However, such $d$-electron superconductivities appear only in materials with specific combinations of transition metal elements and crystal structure. For example, almost all transition metal compounds that are isostructural to the above superconductors but contain other transition metal elements do not exhibit superconductivity. A recently discovered layered nickelate superconductor, a Ni analogue of high-$T_c$ cuprates, is a rare counter example[9]. This nature of $d$-electron superconductors prevents a complete understanding of their features based on systematic experimental studies.

Here, we report the ternary telluride series Sc$_6$$M$Te$_2$ as a unique family of $d$-electron superconductors incorporating 3$d$, 4$d$, and 5$d$ electron systems. Sc$_6$$M$Te$_2$ compounds with $M$ = Mn, Fe, Co, Ni, Ru, Rh, Os, and Ir have been synthesized and reported to crystallize in the hexagonal Zr$_6$CoAl$_2$-type structure with space group $P$–62$m$ without inversion symmetry, but their physical properties have not been reported thus far[10,11]. A characteristic point of this crystal structure is the fact that the $M$ atoms are trigonal prismatically coordinated by six Sc atoms and form one-dimensional chains along the $c$ axis, as shown in Fig. 1(a). This crystal structure is one of the ordered Fe$_2$P-types, in which the P sites are regularly occupied by $M$ and Te atoms in a 1:2 ratio. In terms of superconductivity, transition metal compounds with ZrNiAl-type structure, which is another ordered Fe$_2$P-type in which the Fe sites are regularly occupied by two kinds of metal atoms, have been studied as 4$d$- and 5$d$-electron superconductors, as represented by ZrRuP[12-15].

Sc$_6$$M$Te$_2$ ($M$ = Mn, Fe, Co, Ni, Ru, Rh, Os, and Ir) polycrystalline samples, as shown in Fig. 1(b) for $M$ = Fe, were synthesized by the arc-melting method (See Supplementary Note 1). Powder X-ray diffraction (XRD) data measured by employing Cu K$\alpha$ radiation at room temperature using a MiniFlex diffractometer (RIGAKU), shown in Supplementary Fig. 1, confirmed that Sc$_6$$M$Te$_2$ with Zr$_6$CoAl$_2$-type crystal structure, reported in previous studies[10,11], was obtained as the main phase. For $M$ = Fe, Co, Ni, Ru, and Ir, the obtained samples consisted of a single phase of Sc$_6$$M$Te$_2$. Electrical resistivity and heat capacity measurements were performed using a Physical Property Measurement System (Quantum Design). Electrical resistivity measurements down to 0.1 K were performed using an adiabatic demagnetization refrigerator. Electrical resistivity and heat capacity measurements down to 0.5 K under magnetic fields were performed using a $^3$He refrigerator. Magnetization was measured using a Magnetic Property Measurement System (Quantum Design). Electronic structure calculations were performed using the WIEN2k package (See

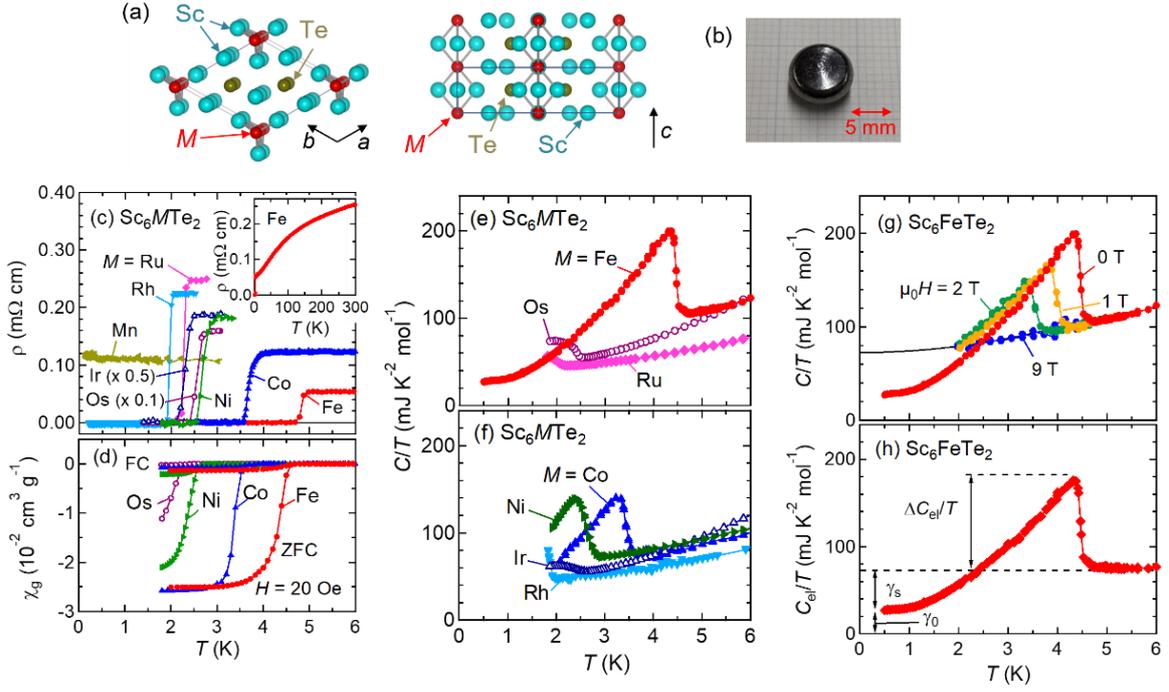

Figure 1. (a) Crystal structure of Sc$_6$$M$Te$_2$ viewed along (left) and perpendicular to the $c$ axis (right) drawn using VESTA[17]. The solid line indicates the unit cell. (b) Sc$_6$FeTe$_2$ polycrystalline sample synthesized by the arc-melting method. (c) Temperature dependence of electrical resistivity of Sc$_6$$M$Te$_2$ polycrystalline samples. The inset shows the Sc$_6$FeTe$_2$ data at high temperatures. (d) Temperature dependence of field-cooled (FC) and zero-field cooled (ZFC) magnetic susceptibility of Sc$_6$$M$Te$_2$ polycrystalline samples measured at a magnetic field of 20 Oe. (e, f) Temperature dependence of heat capacity divided by temperature, $C/T$, for Sc$_6$$M$Te$_2$ polycrystalline samples. (g) Temperature dependence of heat capacity divided by temperature, $C/T$, of Sc$_6$FeTe$_2$ measured at various magnetic fields from 0 to 9 T. The solid curve shows a fit to the equation $C/T = AT^2 + \gamma$ of the 1.9–3.5 K data taken at $\mu_0H$ = 9 T, yielding $A$ = 1.6(2) mJ K$^{-4}$ mol$^{-1}$ and $\gamma$ = 73(1) mJ K$^{-2}$ mol$^{-1}$. (h) Electron heat capacity divided by temperature, $C_{el}/T$, for Sc$_6$FeTe$_2$ obtained by subtracting the lattice contribution from the data shown in (g). The broken line shows the electron heat capacity in the normal state $\gamma$.

Supplementary Note 2)[16].

The temperature dependence of electrical resistivity, r, of a Sc$_6$FeTe$_2$ polycrystalline sample is shown in Fig. 1(c). As shown in the inset, Sc$_6$FeTe$_2$ showed metallic ρ, decreasing on lowering the temperature. When the temperature was further decreased, r showed a sharp drop to zero between 5.0 and 4.7 K. As shown in Fig. 1(d), the zero-field-cooled (ZFC) and field-cooled (FC) magnetization data showed strong diamagnetic signals below 4.7 K, indicating that a bulk superconducting transition occurred at this temperature. The shielding fraction at 1.8 K was estimated to be high, at 150%, considerably larger than 100%, probably due to a demagnetization effect. The jump observed in the heat capacity data shown in Fig. 1(e) also supported the emergence of bulk superconductivity. Taking the zero-resistivity temperature, the onset of the magnetization drop, and the onset of the heat capacity jump to be 4.7 K, the $T_c$ was determined as 4.7 K.

Bulk superconducting transitions also appeared in the other Sc$_6$$M$Te$_2$ polycrystalline samples, except for $M$ = Mn. Their ρ data showed zero resistivity, as shown in Fig. 1(c). Corresponding to these superconducting transitions, large diamagnetic signals and jumps in heat capacity appeared in the magnetization and heat capacity data, as shown in Fig. 1(d)

Table I. $T_c$ and $H_{c2}(0)$ of Sc$_6$$M$Te$_2$. The $T_c$ values were determined from the zero-resistivity temperature, the onset of the magnetization drop, and the onset of the heat capacity jump. The $H_{c2}(0)$ and $\xi_{GL}$ values were determined by fitting the midpoint of the resistivity drop $T_c^{mid}$ to the GL formula.

| 3$d$ element | Mn | Fe | Co | Ni |
|---|---|---|---|---|
| $T_c$ (K) | < 0.1 | 4.7 | 3.6 | 2.7 |
| $\mu_0H_{c2}(0)$ (T) | | 8.68(8) | 6.19(6) | 2.61(7) |
| $\xi_{GL}$ (nm) | | 6.16(3) | 7.29(3) | 11.2(2) |
| 4$d$ element | | Ru | Rh | |
| $T_c$ (K) | | 1.9 | 1.9 | |
| $\mu_0H_{c2}(0)$ (T) | | 3.55(5) | 3.52(5) | |
| $\xi_{GL}$ (nm) | | 9.63(7) | 9.66(7) | |
| 5$d$ element | | Os | Ir | |
| $T_c$ (K) | | 2.4 | 2.0 | |
| $\mu_0H_{c2}(0)$ (T) | | 5.23(3) | 5.00(4) | |
| $\xi_{GL}$ (nm) | | 7.93(2) | 8.11(3) | |

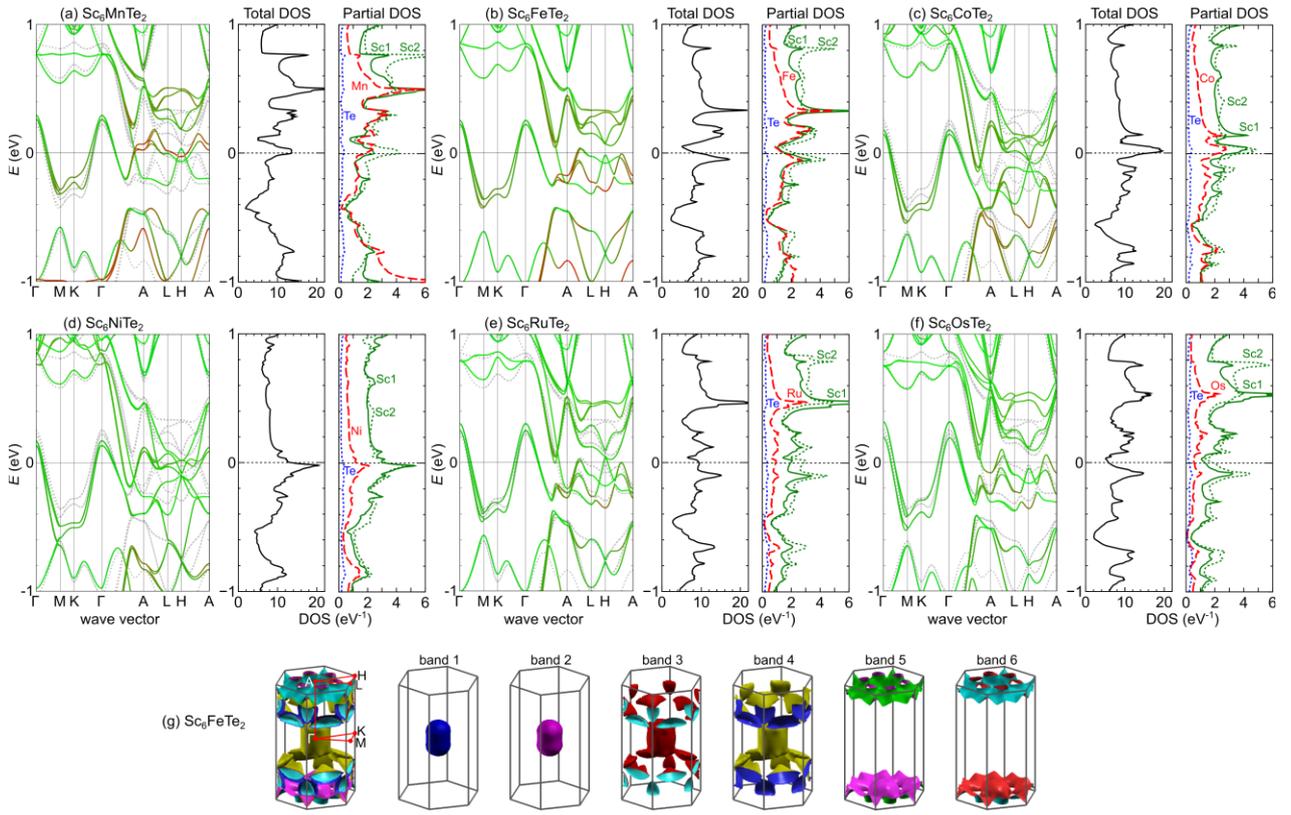

Figure 2. Electronic states of Sc$_6M$Te$_2$ with spin–orbit coupling. (a–f) Electronic structures (left) and DOS (right) of Sc$_6M$Te$_2$ for $M$ = Mn (a), Fe (b), Co (c), Ni (d), Ru (e), and Os (f). The green and red colors in the electronic structures represent the contributions of Sc 3$d$ and $M$ 3$d$/4$d$/5$d$ orbitals, respectively. The Fermi level is set to 0 eV. The dotted lines are the band dispersions for $M$ = Fe. The partial DOS in each figure shows the contributions from three Sc atoms occupying Sc1 site ($z$ = 0.5, 3g Wyckoff position), three Sc atoms occupying Sc2 site ($z$ = 0, 3f Wyckoff position), one $M$ atom, and two Te atoms in the primitive unit cell. (g) The Fermi surfaces of Sc$_6$FeTe$_2$ drawn using XCrySDen Program[18].

and Figs. 1(e, f), respectively, indicating the emergence of bulk superconductivity. As shown in Table 1, seven compounds, except for Sc$_6$MnTe$_2$, exhibited a bulk superconducting transition above 1.8 K. Sc$_6$MnTe$_2$ did not show superconductivity above 0.1 K.

The $T_c$ values of Sc$_6M$Te$_2$ showed characteristic $M$ dependences. Four compounds with $M$ = 4$d$ and 5$d$ elements displayed almost the same $T_c$ of approximately 2 K, as shown in Table 1, but those with $M$ = 3$d$ elements displayed higher values and increased in the order of Ni, Co, and Fe. Sc$_6$NiTe$_2$ showed $T_c$ = 2.7 K, which increased with decreasing atomic number of $M$, and Sc$_6$FeTe$_2$ showed the highest $T_c$ of 4.7 K.

In the following, we discuss the role of $M$ atoms in the chemical trend of $T_c$ in Sc$_6M$Te$_2$. First-principles calculations have shown that the 3$d$ orbitals of $M$ atoms are able to make a large contribution to the electronic states near the Fermi energy, $E_F$. As shown in Fig. 2(b), $E_F$ of Sc$_6$FeTe$_2$ is located just above the peak of the density of states, $D(E)$, and the contributions from Fe and Sc, which mainly involve their 3$d$ orbitals, are dominant for $D(E_F)$. A similar tendency was seen for compounds with other 3$d$ elements, but the contribution from $M$ 3$d$ orbitals becomes smaller with increasing atomic number, as shown in Figs. 2(a–d). The contribution of Fe and Mn 3$d$ orbitals in Sc$_6$FeTe$_2$ and Sc$_6$MnTe$_2$ are much larger than that of Ni 3$d$ orbitals in Sc$_6$NiTe$_2$. In contrast, for Sc$_6$RuTe$_2$ and Sc$_6$OsTe$_2$, as shown in Figs. 2(e) and 2(f), respectively, mainly Sc contributes to the electronic state at $E_F$ and the contributions from Ru and Os are considerably smaller than the Fe contribution in Sc$_6$FeTe$_2$. These chemical trends are natural, considering the energy levels of the $d$ orbitals for each $M$. As shown in Supplementary Fig. 2, the energy levels of the 3$d$ orbitals become lower in the order of Mn, Fe, Co, and Ni, and those of the Ru 4$d$ and Os 5$d$ orbitals are also lower than those of the Fe 3$d$ orbitals. A point to be mentioned in this figure is more flat 3$d$ bands in $M$ = Ni than in those of Mn. This suggests that the stronger hybridization between $M$ and ligand orbitals occurs in Sc$_6M$Te$_2$, contrary to the case of oxides, because $M$ atoms are coordinated by more positive Sc atoms.

As shown in Fig. 2(g), Sc$_6$FeTe$_2$ has two kinds of Fermi surfaces. Firstly, there are spherical hole surfaces surrounding the Γ point, which are slightly elongated along the $k_z$ direction. Secondly, there are complex-shaped surfaces, which mainly appear at $\pi/c > |k_z| > \pi/2c$ of the Brillouin zone. The former surfaces mainly consist of Sc 3$d$ orbitals, and are a common feature of all of the studied $M$, as shown in Figs. 2(a–f). In contrast, the latter surfaces have considerable contributions from $M$ $3d$ orbitals for $M$ = Mn, Fe, and Co, in addition to the Sc $3d$ orbitals, in contrast to the Ru and Os cases, in which contributions from the Ru/Os $4d/5d$ orbitals are small. The very similar $T_c$ values for $M$ = $4d$ and $5d$ elements suggest that the superconductivity at approximately 2 K appears when the Sc $3d$ electrons dominantly contribute to the electronic state at $E_F$. In contrast, the higher and element-dependent $T_c$ for $M$ = $3d$ elements indicates that the latter Fermi surfaces, with a significant contribution from the $M$ $3d$ orbitals, contribute to the emergence of superconductivity with higher $T_c$.

The question then arises as to how the $3d$ electrons of $M$ atoms contribute to the superconductivity in Sc$_6M$Te$_2$. The $D(E_F)$ of 9.5 eV$^{-1}$ for Sc$_6$RuTe$_2$ ($T_c$ = 1.9 K) is almost same as that of 9.3 eV$^{-1}$ for Sc$_6$FeTe$_2$ ($T_c$ = 4.7 K), as shown in Figs. 2(b) and 2(e), respectively. The Sommerfeld coefficients calculated from these $D(E_F)$ values are $\gamma_{band}$ = 22.0 and 22.3 mJ K$^{-2}$ mol$^{-1}$ for Fe and Ru, respectively. In the case of Ru, the experimentally determined Sommerfeld coefficient is $\gamma$ = 40 mJ K$^{-2}$ mol$^{-1}$, which is almost twofold larger than $\gamma_{band}$ (Supplementary Fig. 4). In contrast, $\gamma$ = 73 mJ K$^{-2}$ mol$^{-1}$ for Sc$_6$FeTe$_2$ [Fig. 1(g)] is more than three times larger than $\gamma_{band}$, suggesting that the $\gamma$ of Sc$_6$FeTe$_2$ is more significantly enhanced by strong electron correlation and/or electron-phonon interaction of the Fe $3d$ electrons. Moreover, the experimental $\gamma$ for Fe is larger than those for Co and Ni, which is natural considering the contribution of $3d$ orbitals at $E_F$ discussed above ($\gamma$ = 55 mJ K$^{-2}$ mol$^{-1}$ for Co and 60 mJ K$^{-2}$ mol$^{-1}$ for Ni; see Supplementary Fig. 4) and there is a positive correlation between the experimental $\gamma$ and $T_c$ values in Sc$_6M$Te$_2$, as shown in Supplementary Fig. 5. These results strongly suggest that the $3d$ electrons of Fe play a more significant role in realizing higher $T_c$ than those in other Sc$_6M$Te$_2$ systems.

As shown in Fig. 2(d), the electron heat capacity divided by temperature, $C_{el}/T$, for $M$ = Fe exhibited a large jump associated with the superconducting transition and saturating behavior with $\gamma_0$ = 28 mJ K$^{-2}$ mol$^{-1}$ toward $T$ = 0. At present, the origin of $\gamma_0$ is unclear. There is a possibility that the paramagnetic phase remains at $T$ = 0 due to its extrinsic origin, such as disorder and defects in the sample. In this case, taking $\Delta C_{el}/T_c$ = 108 mJ K$^{-2}$ mol$^{-1}$ and $\gamma - \gamma_0$ = 45 mJ K$^{-2}$ mol$^{-1}$, the magnitude of the jump at $T_c$ was estimated as $\Delta C_{el}/(\gamma - \gamma_0)T_c$ = 2.40. This value greatly exceeds the weak-coupling limit of 1.43 for conventional Bardeen–Cooper–Schrieffer (BCS) superconductivity, suggestive of the strong coupling superconductivity. Another possibility is that the saturating behavior at $\gamma_0$ is intrinsic possibly due to multigap formation. It would be interesting if the spin and orbital fluctuations of $3d$ electrons in Fe atoms are to contribute to these possible scenarios. In fact, in the case of $M$ = Mn, next to Fe in the periodic table, superconductivity was not observed above 0.1 K. As shown in Supplementary Fig. 6, the magnetic susceptibility of Sc$_6$MnTe$_2$ strongly increases below 50 K, unlike that of Sc$_6$CoTe$_2$, showing Pauli paramagnetic behavior. It is most likely that Cooper pair formation is inhibited by the strong magnetism of Mn $3d$ electrons.

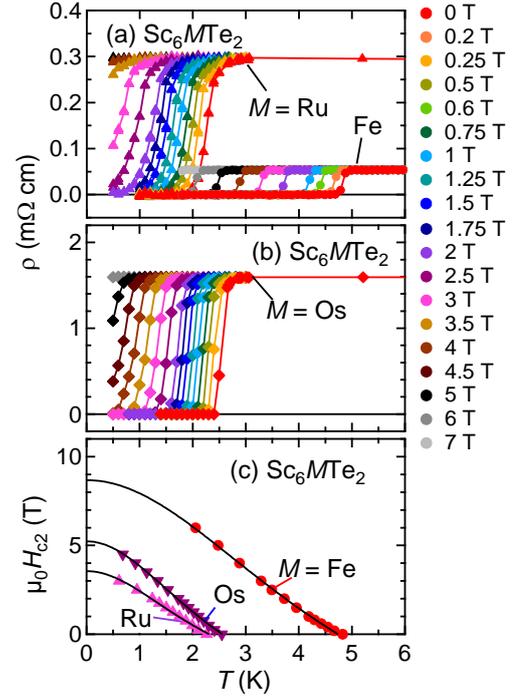

Figure 3. The electrical resistivity of Sc$_6M$Te$_2$ under various magnetic fields. (a, b) Temperature dependences of electrical resistivity of Sc$_6M$Te$_2$ polycrystalline samples with $M$ = Fe, Ru (a), and Os (b), measured at various magnetic fields. (c) Temperature dependences of the upper critical field $H_{c2}(T)$ determined from the midpoint of the drop $T_c^{mid}$ in the resistivity data. The solid curves show fits to the GL formula.

The upper critical field $H_{c2}(T)$ is also strongly dependent on the $M$ elements in Sc$_6M$Te$_2$. Figure 3 and Supplementary Fig. 7 show the $\rho$ of Sc$_6M$Te$_2$ measured at various magnetic fields and $H_{c2}(T)$ determined from the midpoint of the resistivity drop $T_c^{mid}$ at each magnetic field, respectively. The $H_{c2}$ data were fitted to the Ginzburg–Landau (GL) formula $H_{c2}(T) = H_{c2}(0)[1 - (T/T_c^{mid})^2]/[1 + (T/T_c^{mid})^2]$, yielding the $\mu_0 H_{c2}(0)$ and GL coherence length $\xi_{GL}$ values shown in Table 1. The value of $T_c^{mid}$ at each magnetic field was taken as the superconducting transition temperature in this analysis, because the zero-resistivity temperature in magnetic fields might be underestimated due to the stronger resistivity tails at around zero-resistivity temperature compared to those in zero magnetic field. As can be seen from the slopes of the fitting curves in Fig. 3, the ratio of $H_{c2}(0)$ to $T_c^{mid}$ for $M$ = Os is larger than those for $M$ = Fe and Ru. Thus, $\mu_0 H_{c2}(0)$ = 5.23 for $M$ = Os exceed the Pauli limit determined by paramagnetic effects in the singlet superconducting states, $\mu_0 H_{c2}(0)/T$ = 1.84$T_c^{mid}$/K = 4.69[19]. The similar situation is realized for $M$ = Ir (Supplementary Fig. 7). In contrast, the $H_{c2}(0)$ values for the other five $M$ cases are lower than the Pauli limit. These results suggest that the high $H_{c2}(0)$ values for $M$ = Os and Ir originate from the $5d$ electrons of the $M$ atoms.

There are two possible mechanisms that might explain the emergence of high $H_{c2}(0)$ in Sc$_6M$Te$_2$ with $M$ = 5$d$ elements. One is the contribution of triplet Cooper pairs in the parity-mixing superconductivity, as discussed in the noncentrosymmetric heavy-fermion superconductor CePt$_3$Si[20, 21]. The other possibility is weakening of the paramagnetic effect due to strong spin–orbit coupling, as is operative in Chevrel-phase PbMo$_6$S$_8$[22,23]. In both mechanisms, strong spin–orbit coupling of $M$ 5$d$ electrons is essential, demonstrating that the characteristic features of $d$-electrons of $M$ can clearly be invoked in the elemental dependence of $H_{c2}$, similar to that of $T_c$ discussed above. However, the violations of the Pauli limit in Sc$_6$OsTe$_2$ and Sc$_6$IrTe$_2$ are not significant, probably due to the small contribution of the 5$d$ orbitals to the electronic state at $E_F$. It is expected that the effect of strong spin–orbit coupling will be observed more clearly by discovering new members with stronger contributions of 5$d$ electrons at $E_F$.

In summary, the Sc$_6M$Te$_2$ series was found to comprise unique $d$-electron superconductors that exhibit bulk superconductivity for all $M$ = 3$d$, 4$d$, and 5$d$ elements. The critical temperature is approximately 2 K for $M$ = 4$d$ and 5$d$ elements, but higher for $M$ = 3$d$ elements. The $T_c$ increases in the order of $M$ = Ni, Co, and Fe, and the superconductivity in Sc$_6$FeTe$_2$ has the highest $T_c$ of 4.7 K, which is consistent with the chemical trend of the first principles calculations data. The upper critical field also shows a pronounced element dependence between 3$d$, 4$d$, and 5$d$ elements. These results clearly show that the Sc$_6M$Te$_2$ series constitutes a unique superconductor family, in which $d$ electrons of 3$d$, 4$d$, and 5$d$ $M$ atoms strongly influence the superconducting properties.

**Acknowledgments**

The authors are grateful to R. Ishii for her support in sample preparation, T. Yamauchi for his help in magnetization measurements, and Y. Shimizu, K. Doi, H. Takei, S. Nakata, H. Wadati, K. Yuchi, and Z. Hiroi for helpful discussion. This work was supported by the Collaborative Research Project of Materials and Structures Laboratory, Tokyo Institute of Technology and JSPS KAKENHI (Grant Nos. 18H04314, 19H05823, 20H02603, 23H01831).

*E-mail: yokamoto@issp.u-tokyo.ac.jp

**Supplementary Note 1. Sample preparation and characterization of polycrystalline samples of Sc$_6M$Te$_2$**

Sc$_6M$Te$_2$ ($M$ = Mn, Fe, Co, Ni, Ru, Rh, Os, and Ir) polycrystalline samples were synthesized by the arc-melting of Sc chips (99%, Kojundo Chemical Lab.) and $M$ and Te powders (99.99%, RARE METALLIC)[11]. Mn (99.9%, Kojundo Chemical Lab.), Fe (99.9%, Kojundo Chemical Lab.), Co (99%, Kojundo Chemical Lab.), Ni (99.9%, Kojundo Chemical Lab.), Ru (99.95%, RARE METALLIC), Rh (99.95%, RARE METALLIC), Os (99.9%, RARE METALLIC), and Ir (99.9%, Kojundo Chemical Lab.) powders were used for $M$. Firstly, Sc chips, $M$ powder, and Te powder were weighed in a 6:1:2 molar ratio. The $M$ and Te powders were then mixed and pressed into a pellet. The Sc chips and the pellet were placed on a water-cooled copper hearth and arc-melted under Ar atmosphere. The obtained buttons were subsequently inverted and arc-melted several times more to promote homogenization. For $Sc_6NiTe_2$, the obtained button was annealed in an evacuated quartz tube at 1273 K for 6 h and then cooled to room temperature over 120 h.

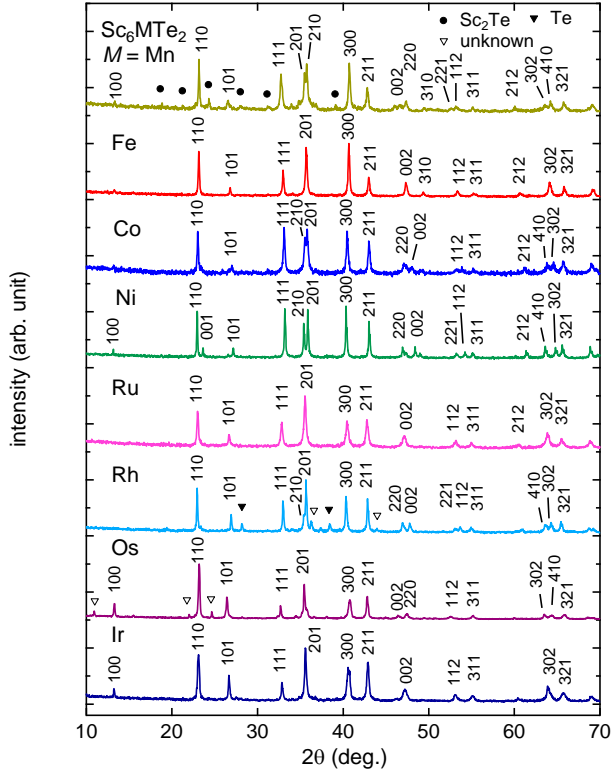

Supplementary Figure 1. Powder XRD patterns of polycrystalline samples of $Sc_6MTe_2$ ($M$ = Mn, Fe, Co, Ni, Ru, Rh, Os, and Ir) measured at room temperature. Filled circles and filled triangles indicate the diffraction peaks of $Sc_2Te$ and Te impurity phases, respectively. Open triangles indicate those of unknown impurities. Peak indices are given hexagonal unit cells with $P$–$62m$ symmetry.

$Sc_6MTe_2$ ($M$ = Mn, Fe, Co, Ni, Ru, Rh, Os, and Ir) polycrystalline samples were characterized by powder X-ray diffraction (XRD) analysis employing Cu Kα radiation at room temperature using a MiniFlex diffractometer (RIGAKU). As shown in Supplementary Fig. 1, all of the peaks, except for small peaks from impurity phases, $Sc_2Te$, Te, and unknown impurities, could be indexed on the basis of a hexagonal unit cell with $P$–$62m$ symmetry with the lattice constants shown in Supplementary Table 1. The obtained lattice constants are almost the same as those reported in previous studies[10,11]. These results indicated that $Sc_6MTe_2$ with $Zr_6CoAl_2$-type crystal structure was obtained as the main phase. For $M$ = Fe, Co, Ni, Ru, and Ir, the obtained samples consisted of a single phase of $Sc_6MTe_2$.

Supplementary Table 1. Lattice constants (Å) and cell volumes (Å$^3$) at room temperature for $Sc_6MTe_2$ ($M$ = Mn, Fe, Co, Ni, Ru, Rh, Os, and Ir).

| compound | $a$ | $c$ | $V$ |
| --- | --- | --- | --- |
| $Sc_6MnTe_2$ | 7.673(3) | 3.900(3) | 198.9(2) |
| $Sc_6FeTe_2$ | 7.683(2) | 3.838(3) | 196.2(2) |
| $Sc_6CoTe_2$ | 7.718(2) | 3.780(3) | 195.0(2) |
| $Sc_6NiTe_2$ | 7.741(2) | 3.764(2) | 195.3(1) |
| $Sc_6RuTe_2$ | 7.719(4) | 3.855(5) | 198.9(3) |
| $Sc_6RhTe_2$ | 7.740(2) | 3.808(2) | 197.6(1) |
| $Sc_6OsTe_2$ | 7.660(2) | 3.915(2) | 198.9(1) |
| $Sc_6IrTe_2$ | 7.691(2) | 3.861(2) | 197.8(1) |

**Supplementary Note 2. Details on the first principles calculations**

Electronic structure calculations were performed using the WIEN2k package[16] in the framework of density functional theory (DFT) based on the full-potential linearized augmented plane wave (FP-LAPW) method. We used the Perdew–Burke–Ernzerhof (PBE) functional, which is one of the most widely used generalized gradient approximation (GGA) based exchange-correlation functionals. The core separation energy cut-off was set to −6.0 Ry, and the maximum modulus of reciprocal vector $K_{max}$ was chosen to satisfy $R_{MT}K_{max}$ = 7.0. In the Brillouin zone, $k$-meshes were set to 16×16×28 for self-consistent DFT calculations and 40×40×60 for calculations of the density of states and Fermi surfaces. Experimentally obtained structural parameters were used for $Sc_6FeTe_2$, $Sc_6NiTe_2$, and $Sc_6OsTe_2$[10,11]. For $Sc_6MnTe_2$, $Sc_6CoTe_2$, and $Sc_6RuTe_2$, experimentally obtained lattice constants were used in combination with the $x$ coordinates of Sc sites for $M$ = Fe, Ni, and Rh, respectively[10,11]. The Fermi surfaces were drawn using XCrySDen program[18].

**Supplementary Note 3. Electronic States of $Sc_6MTe_2$**

Supplementary Fig. 2 shows the electronic structures and DOS of $Sc_6MTe_2$ ($M$ = Mn, Fe, Co, Ni, Ru, Os) calculated with spin–orbit coupling. Figures (a–f) in the main text are the enlarged view of this figure. For $Sc_6MnTe_2$ and $Sc_6FeTe_2$, Sc and Fe contributions are dominant for $D(E_F)$. However, the M contribution becomes smaller in $Sc_6CoTe_2$ and $Sc_6NiTe_2$, corresponding to the fact that the energy levels of the 3$d$ orbitals become lower in the order of Mn, Fe, Co, and Ni. The Te contributions for $D(E_F)$ is small for all cases.

Supplementary Fig. 3 shows Fermi surfaces of $Sc_6MTe_2$ ($M$ = Mn, Co, Ni, Ru, and Os) calculated with spin–orbit coupling. Fermi surfaces of $Sc_6FeTe_2$ are shown in Fig. 3(g) in the main text. As shown in these figures, $Sc_6MTe_2$ commonly has hole pockets surrounding the G point. $Sc_6MTe_2$

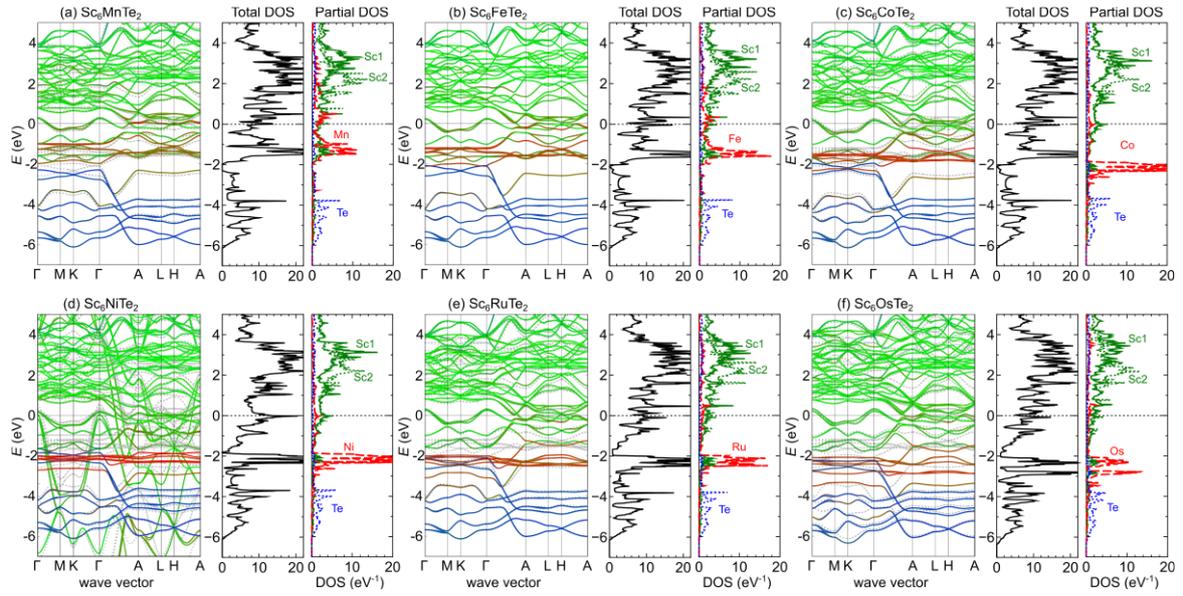

Supplementary Figure 2. Electronic structures (left) and DOS (right) of Sc$_6M$Te$_2$ for $M$ = Mn (a), Fe (b), Co (c), Ni (d), Ru (e), and Os (f), calculated using the WIEN2k package[16]. The green and red colors in the electronic structures represent the contributions of Sc 3$d$ and $M$ 3$d$/4$d$/5$d$ orbitals, respectively. The Fermi level is set to 0 eV. The dotted lines are the band dispersions for $M$ = Fe. The partial DOS in each figure shows the contributions from three Sc atoms occupying Sc1 site ($z$ = 0.5, 3g Wyckoff position), three Sc atoms occupying Sc2 site ($z$ = 0, 3f Wyckoff position), one $M$ atom, and two Te atoms in the primitive unit cell.

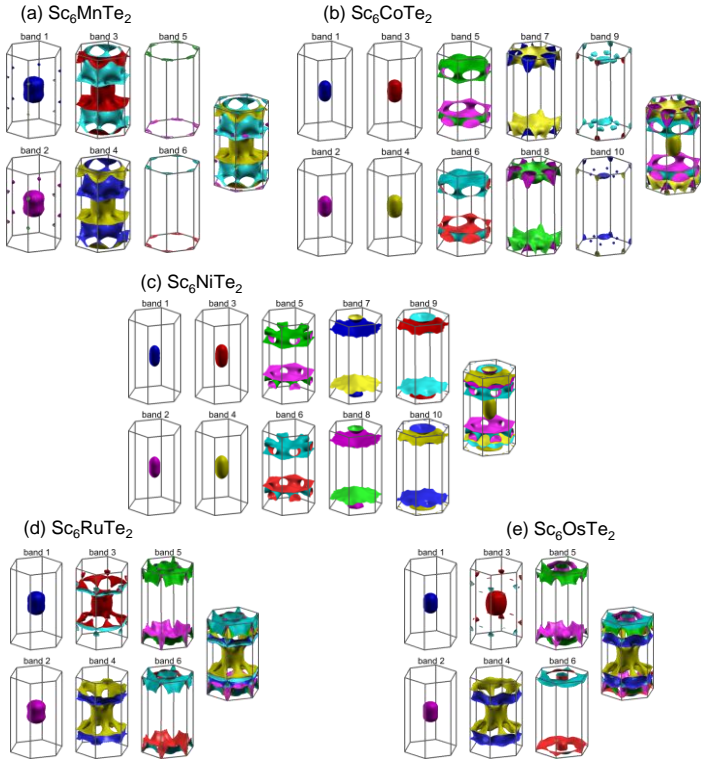

Supplementary Figure 3. Fermi surfaces of Sc$_6M$Te$_2$ ($M$ = Mn, Co, Ni, Ru, and Os) calculated with spin-orbit coupling. They were drawn using XCrySDen program[18].

also has complex-shaped surfaces, mainly existing at $\pi/c > |k_z| > \pi/2c$ of the Brillouin zone, the shapes of which depend on the $M$ element.

**Supplementary Note 4. Estimation of Sommerfeld coefficient of Sc$_6M$Te$_2$**

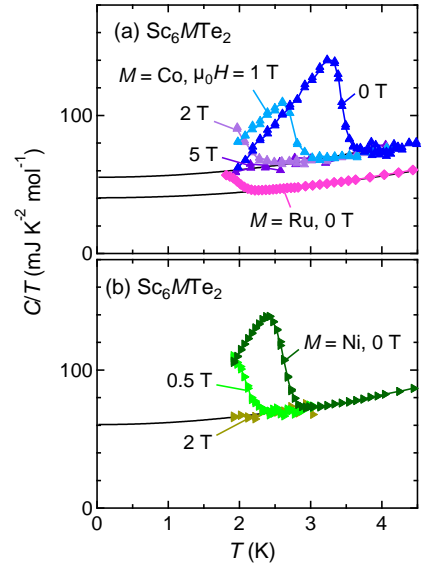

Supplementary Figure 4. Temperature dependence of heat capacity divided by temperature, $C/T$, for Sc$_6M$Te$_2$ ($M$ = Co, Ni, and Ru) polycrystalline samples. The data for Co and Ru are shown in (a), and those for Ni are shown in (b). The solid curves show fits to the equation $C/T = AT^2 + \gamma$.

Supplementary Fig. 4 shows the temperature dependence of heat capacity divided by temperature, $C/T$, for Sc$_6$CoTe$_2$, Sc$_6$NiTe$_2$, and Sc$_6$RuTe$_2$ polycrystalline samples. The 5 T data for $M$ = Co (1.9-4.0 K), 2 T data for Ni (1.9-3 K), and 0 T data for Ru (2.4–3.4 K) were fitted to the equation $C/T = AT^2 + \gamma$, yielding Sommerfeld coefficients of $\gamma$ = 55(1), 60(1), and 40.4(1) mJ K$^{-2}$ mol$^{-1}$ and $A$ = 1.4(1), 1.5(2), and 0.99(1) mJ K$^{-4}$ mol$^{-1}$, respectively. Supplementary Fig. 5 shows the critical temperature versus Sommerfeld coefficient in Sc$_6$$M$Te$_2$. There is a positive correlation between $T_c$ and $\gamma$.

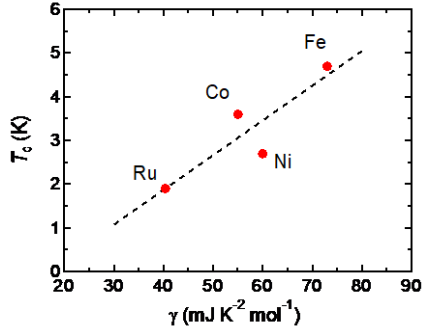

Supplementary Figure 5. The critical temperature versus Sommerfeld coefficient in Sc$_6$$M$Te$_2$.

**Supplementary Note 5. Magnetic susceptibility in the normal state of Sc$_6$MnTe$_2$ and Sc$_6$CoTe$_2$**

As shown in Supplementary Fig. 6, magnetic susceptibility, $\chi_{mol}$, of Sc$_6$CoTe$_2$ showed Pauli paramagnetic behavior, being almost independent of temperature. The sudden decrease of $\chi_{mol}$ at low temperature is due to the superconducting transition. In contrast, $\chi_{mol}$ of Sc$_6$MnTe$_2$ strongly increases below 50 K especially for the 0.1 T data. Because there are no differences between the zero-field-cooled and field-cooled data, this increase is not due to magnetic long-range order, but strong magnetism, probably arising from the Mn atoms in Sc$_6$MnTe$_2$, which might prevent Cooper pair formation therein.

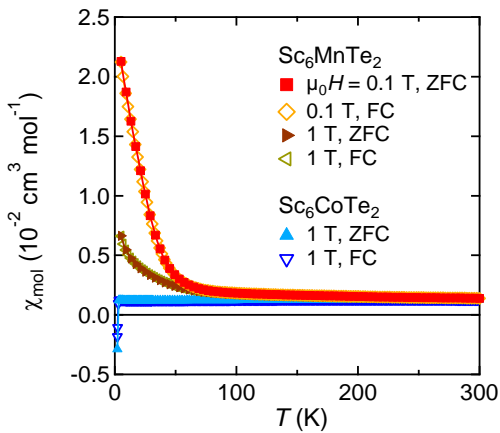

Supplementary Figure 6. Temperature dependences of zero-field-cooled (ZFC) and field-cooled (FC) magnetic susceptibilities of Sc$_6$MnTe$_2$ and Sc$_6$CoTe$_2$ polycrystalline samples. The data were taken at magnetic fields of 0.1 and 1 T for Sc$_6$MnTe$_2$ and at a magnetic field of 1 T for Sc$_6$CoTe$_2$.

**Supplementary Note 6. Electrical resistivity of Sc$_6$$M$Te$_2$ under magnetic fields**

Supplementary Figs. 7(a,b) show the electrical resistivity of Sc$_6$$M$Te$_2$ ($M$ = Co, Rh, Ir, and Ni) measured at various magnetic fields. The upper critical field $H_{c2}(T)$ determined from the midpoint of the resistivity drop at each magnetic field for these compounds is shown in Supplementary Fig. 6(c). The $H_{c2}$ data are fitted to the Ginzburg–Landau (GL) formula $H_{c2}(T) = H_{c2}(0)[1 - (T/T_c)^2]/[1 + (T/T_c)^2]$, yielding the $\mu_0 H_{c2}(0)$ and GL coherence lengths $\xi_{GL}$ values shown in Table 1.

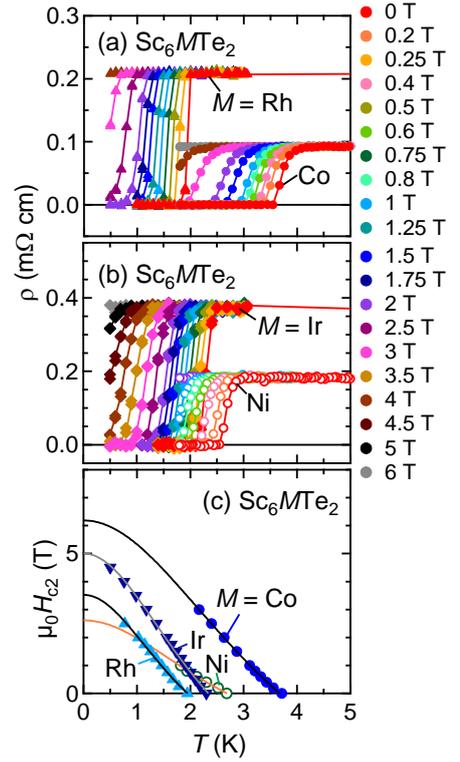

Supplementary Figure 7. The electrical resistivity of Sc$_6$$M$Te$_2$ under various magnetic fields. (a, b) Temperature dependences of electrical resistivity of Sc$_6$$M$Te$_2$ polycrystalline samples with $M$ = Co, Rh (a), Ir, and Ni (b), measured at various magnetic fields. (c) Temperature dependences of the upper critical field $H_{c2}(T)$ determined from the midpoint of the drop $T_c^{mid}$ in the resistivity data. The solid curves show fits to the GL formula.